\documentclass[prl,twocolumn,showpacs,preprintnumbers,amsmath,amssymb]{revtex4}
\usepackage{graphicx}
\usepackage{bm}

\begin{document}

\preprint{ }
\title{Quantum oscillations of rectified dc voltage as a
function of magnetic field in an "almost" symmetric
superconducting ring}


\author{V.~I. Kuznetsov}
\email{kvi@ipmt-hpm.ac.ru}
\author{A.~A. Firsov}
\author{S.~V. Dubonos}
\affiliation{Institute of Microelectronics Technology and High
Purity Materials, Russian Academy of Sciences, 142432
Chernogolovka, Moscow Region, Russia}

\date{\today}

\begin{abstract}
Periodic quantum oscillations of a rectified dc voltage
$V_{dc}(B)$ vs the perpendicular magnetic field $B$ were measured
near the critical temperature $T_{c}$ in a single superconducting
aluminum almost symmetric ring (without specially created circular
asymmetry) biased by alternating current with a zero dc component.
With varying bias current and temperature, these $V_{dc}(B)$
oscillations behave like the $V_{dc}(B)$ oscillations observed in
a circular-asymmetric ring but are of smaller amplitude. The
Fourier spectra of the $V_{dc}(B)$ functions exhibit a fundamental
frequency, corresponding to the ring area, and its higher
harmonics. Unexpectedly, satellite frequencies depending on the
structure geometry and external parameters were found next to the
fundamental frequency and around its higher harmonics.
\end{abstract}

\maketitle

The effect of ac voltage rectification in a superconducting
thin-film asymmetric ring with a specially created circular
asymmetry was recently reported \cite{c1}. Unlike previously
proposed rectifiers which contained superconducting loops with
tunnel or point contacts \cite{c2}, the asymmetric structure
proposed in \cite{c1} is the simplest and most efficient ac
voltage rectifier with a high magnetic-field-dependent output
signal.

Time-averaged rectified dc voltage $V_{dc}(B)$ was observed in a
circular-asymmetric ring \cite{c1} threaded by a magnetic flux and
biased by a low-frequency sinusoidal current (without a dc
component) with an amplitude close to the critical one at
temperatures slightly below $T_{c}$. $V_{dc}(B)$ is an odd
function with respect to $B$ and is periodically dependent on $B$
with the period $\Delta B=\Phi_{0}/S$, where $\Phi_{0}$ is the
superconducting magnetic flux quantum and $S$ is the effective
ring area \cite{c1}. The experimental results in \cite{c1} provide
indirect arguments for the $V_{dc}(B)$ voltage in a single
asymmetric ring is directly proportional to the ring circulating
current $I_{R}$. It can then be supposed that $V_{dc}(B)$
functions measured at various parameters can be used instead of
$I_{R}$ functions to describe the quantum state of an asymmetric
structure.

The interest to asymmetric rings  \cite{c1} was also due to that
such but extremely thin-wall rings biased by a microwave current
at $T<0.5T_{c}$ could be used in a novel superconducting flux
qubit with quantum phase-slip centers \cite{c3}.

In this work, we report an experimental investigation of ac
voltage rectification in a ring where circular asymmetry was not
specially created. The rings studied were slightly geometrically
distorted, the distortion arose during structure fabrication and
was about $10\%$ of the ring major wire width. In spite of a weak
circular asymmetry of the rings, the observed rectification effect
was fairly large.
\begin{figure}
\includegraphics[width=1.0\columnwidth]{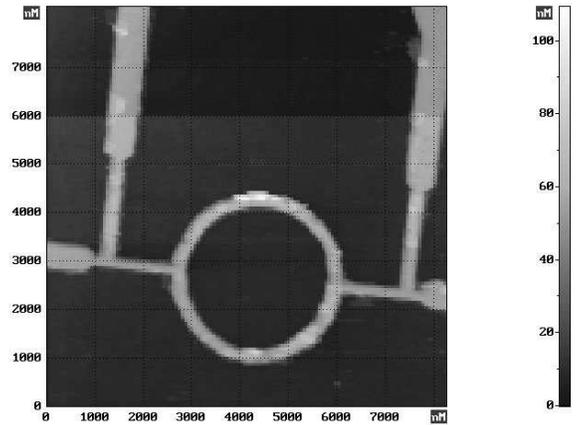}
\caption{\label{image} AFM image of the structure in a $8260
\times 8260$ ${\rm nm}^{2}$ window. The outer diameter of the ring
is $4.1$ $\mu$m.}
\end{figure}

The sample studied was a thin-film aluminum structure $d=50$ nm
thick. It was fabricated by thermal Al deposition onto a silicon
substrate, using the lift-off process of electron beam lithography
with proximity effect correction. An atomic-force microscopy (AFM)
image of the structure in an $8.26 \times 8.26$ $\mu\rm m^{2}$
window is shown in Fig. \ref{image}. The ring with a $10\%$ wire
widening in the upper ring part is in the center of the structure.
The width of all narrow wires, except the widened part, is
$w_{n}=0.43\pm 0.02$ $\mu$m, the width of the widened part is
$w_{w}=0.48\pm 0.02$ $\mu$m (Fig. \ref{image}). The areas of the
inner and outer ring contours are $S_{in}=8.3$ $\mu\rm m^{2}$ and
$S_{en}=13.2$ $\mu\rm m^{2}$, respectively. The averaged
geometrical area of the ring is, then, $S_{g}=10.75$ $\mu\rm
m^{2}$. The structure resistance in the normal state at $T=4.2$ K
is $R_{4.2}=26$ $\Omega$, the ratio of the room to helium
temperature resistance is $R_{300}/R_{4.2}=2$, the electron
effective mean free path is $l=10$ nm, $T_{c}=1.360$ K, and the
superconducting coherence length is $\xi(0)=110$ nm.

\begin{figure}
\includegraphics[width=1.0\columnwidth]{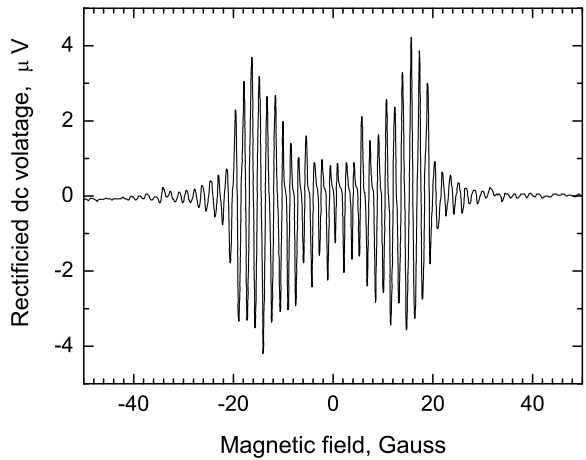}
\caption{\label{voltage} $V_{dc}(B)$ in the structure at a
sinusoidal current of $\nu=1.23$ kHz with $I_{\nu}=2.4$ $\mu$A at
$T=1.335$ K.}
\end{figure}
\begin{figure}
\includegraphics[width=1.0\columnwidth]{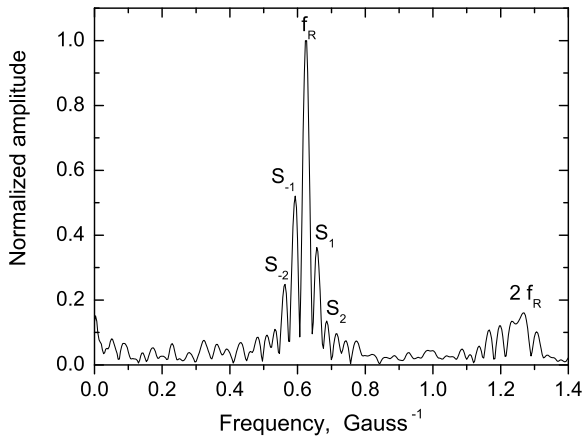}
\caption{\label{fft} Fourier spectrum of the function $V_{dc}(B)$.
The symbols $S_{-2}$, $S_{-1}$, $S_{1}$, and $S_{2}$ refer to
satellite frequencies.}
\end{figure}
$V_{dc}(B)$ oscillations were experimentally studied in structures
similar to that in Fig. \ref{image} at various amplitudes of bias
bias alternating current (without a dc component) close to the
critical amplitude at temperatures slightly lower than $T_{c}$.
The magnetic field perpendicular to the structure surface was
slowly varied during $V_{dc}(B)$ measurements. The $V_{dc}(B)$
curve measured in the structure of Fig. \ref{image} biased by a
sinusoidal current of $\nu=1.23$ kHz with the amplitude
$I_{\nu}=2.4$ $\mu$A at $T=1.335$ K is shown in Fig.
\ref{voltage}. The amplitude of the  $V_{dc}(B)$ oscillations
behaves non-monotonically and is maximum at $17$ Gauss. In fields
higher than 40 Gauss, these oscillations become virtually
imperceptible. As $I_{\nu}$ increases, the highest maximum of the
$V_{dc}(B)$ oscillations shifts towards lower fields, because the
amplitude of the oscillations depends both on bias current and
magnetic field.

For a detailed analysis, the fast Fourier transformation (FFT) of
the $V_{dc}(B)$ curve is calculated. Fig. \ref{fft} shows the FFT
spectrum obtained using $2^{12}$ uniformly distributed points in
the range from $-150$ to $+150$ Gauss. The fundamental frequency
of the ring $f_{R}$ is the inverse value of the fundamental
oscillation period, i.e. $f_{R}=1/\Delta B_{R}=S / \Phi_{0}$. The
fundamental frequency value expected from the ring averaged
geometrical area $S_{g}$ is $f_{g}=0.52$ ${\rm Gauss}^{-1}$.
Indeed, the FFT spectrum exhibits a peak at $f_{R}=0.62$ ${\rm
Gauss}^{-1}$ close to the geometric value $f_{g}$ (Fig.
\ref{fft}).

Apart from the fundamental frequency $f_{R}$, the spectrum
contains its higher harmonics $f_{Rm}=mf_{R}$, where
$m=2,3,4,...$. Moreover, additional satellite peaks around the
fundamental frequency and higher harmonics are observed at
frequencies $f_{Sn}=f_{R}+n \Delta f$ and $f^{m}_{Sn}=mf_{R}+n
\Delta f$ ($n=...,-3,-2,-1,1,2,3,...$), respectively. For this
curve (Fig. \ref{voltage}), the low-frequency value is $\Delta
f=0.03 $ ${\rm Gauss}^{-1}$ and corresponds to the magnetic field
$B_{c}=33$ Gauss. It is seen that $\Delta f$ determines the
low-frequency background, with higher-frequency quantum
oscillations superimposed on it. As external parameters and the
sample geometry change, $B_{c}$ behaves like a field suppressing
the superconducting order parameter in the ring wires but is of
slightly smaller value. So, satellite frequencies arise due to a
combined effect of bias alternating and circulating currents and
are dependent on the magnetic depairing factor.

Thus, rings of geometrical inhomogenecity up to $5\%$ of the major
wire width were fabricated. Some structures had the narrowings
(widenings) by $10\%$ with respect to the wire width. In these
structures, the effect of ac voltage rectification was well
observable. Since the magnetic-field-dependent electron quantum
transport in almost symmetric superconducting loops is usually
studied, using a modulation of the measuring dc current by a weak
bias ac current, the side effect of ac voltage rectification,
which can arise in the structures but was previously disregarded,
should be taken into consideration.

We are grateful to V.~L. Gurtovoi, A.~V. Nikulov, and V.~A. Tulin
for helpful discussions. The work was supported by the program
"Organization of Calculations Based on New Physical Principles"
(Department of Information Technologies and Computer Systems, RAS)
and the program "Quantum Macrophysics" (Presidium of RAS).

\end{document}